\documentstyle[epsf,12pt]{article}

  \def\pp{{\mathchoice
          {
              \kern 1pt%
              \raise 1pt
              \vbox{\hrule width5pt height0.4pt depth0pt
                    \kern -2pt
                    \hbox{\kern 2.3pt
                          \vrule width0.4pt height6pt depth0pt
                          }
                    \kern -2pt
                    \hrule width5pt height0.4pt depth0pt}%
                    \kern 1pt
           }
            {
              \kern 1pt%
              \raise 1pt
              \vbox{\hrule width4.3pt height0.4pt depth0pt
                    \kern -1.8pt
                    \hbox{\kern 1.95pt
                          \vrule width0.4pt height5.4pt depth0pt
                          }
                    \kern -1.8pt
                    \hrule width4.3pt height0.4pt depth0pt}%
                    \kern 1pt
            }
            {
              \kern 0.5pt%
              \raise 1pt
              \vbox{\hrule width4.0pt height0.3pt depth0pt
                    \kern -1.9pt  
                    \hbox{\kern 1.85pt
                          \vrule width0.3pt height5.7pt depth0pt
                          }
                    \kern -1.9pt
                    \hrule width4.0pt height0.3pt depth0pt}%
                    \kern 0.5pt
            }
            {
              \kern 0.5pt%
              \raise 1pt
              \vbox{\hrule width3.6pt height0.3pt depth0pt
                    \kern -1.5pt
                    \hbox{\kern 1.65pt
                          \vrule width0.3pt height4.5pt depth0pt
                          }
                    \kern -1.5pt
                    \hrule width3.6pt height0.3pt depth0pt}%
                    \kern 0.5pt
            }
        }}

  \def\mm{{\mathchoice
                       {
                             \kern 1pt
               \raise 1pt    \vbox{\hrule width5pt height0.4pt depth0pt
                                  \kern 2pt
                                  \hrule width5pt height0.4pt depth0pt}
                             \kern 1pt}
                       {
                            \kern 1pt
               \raise 1pt \vbox{\hrule width4.3pt height0.4pt depth0pt
                                  \kern 1.8pt
                                  \hrule width4.3pt height0.4pt depth0pt}
                             \kern 1pt}
                       {
                            \kern 0.5pt
               \raise 1pt
                            \vbox{\hrule width4.0pt height0.3pt depth0pt
                                  \kern 1.9pt
                                  \hrule width4.0pt height0.3pt depth0pt}
                            \kern 1pt}
                       {
                           \kern 0.5pt
             \raise 1pt  \vbox{\hrule width3.6pt height0.3pt depth0pt
                                  \kern 1.5pt
                                  \hrule width3.6pt height0.3pt depth0pt}
                           \kern 0.5pt}
                       }}

\catcode`@=11
\def\un#1{\relax\ifmmode\@@underline#1\else
        $\@@underline{\hbox{#1}}$\relax\fi}
\catcode`@=12

\let\du=\du                     

\def\a{\alpha}

\def\d{\delta}

\def\f{\phi}

\def\j{\psi}

\def\o{\omega}
\def\p{\pi}
\def\q{\theta}

\def\s{\sigma}
\def\t{\tau}

\def\x{\xi}
\def\z{\zeta}

\def\G{\Gamma}

\def\L{\Lambda}

\def\ve{\varepsilon}

\def\cf{{\cal F}}

\def\ch{{\cal H}}

\def\bo{{\raise-.5ex\hbox{\large$\Box$}}}               
\def\pa{\partial}                                       
\def\pr{\prod}                                          
\def\TH{{\raise.2ex\hbox{$\displaystyle \bigodot$}\mskip-4.7mu \llap H \;}}
\def\face{{\raise.2ex\hbox{$\displaystyle \bigodot$}\mskip-2.2mu \llap {$\ddot
        \smile$}}}                                      

\def\Bar#1{\overline{#1}}                       
\def\VEV#1{\left\langle #1\right\rangle}        
\def\abs#1{\left| #1\right|}                    
\def\leftrightarrowfill{$\mathsurround=0pt \mathord\leftarrow \mkern-6mu
        \cleaders\hbox{$\mkern-2mu \mathord- \mkern-2mu$}\hfill
        \mkern-6mu \mathord\rightarrow$}
\def\dvec#1{\vbox{\ialign{##\crcr
        \leftrightarrowfill\crcr\noalign{\kern-1pt\nointerlineskip}
        $\hfil\displaystyle{#1}\hfil$\crcr}}}           
\def\dt#1{{\buildrel {\hbox{\LARGE .}} \over {#1}}}     

\def\frac#1#2{{\textstyle{#1\over\vphantom2\smash{\raise.20ex
        \hbox{$\scriptstyle{#2}$}}}}}                   
\def\sfrac#1#2{{\vphantom1\smash{\lower.5ex\hbox{\small$#1$}}\over
        \vphantom1\smash{\raise.4ex\hbox{\small$#2$}}}} 
\def\bfrac#1#2{{\vphantom1\smash{\lower.5ex\hbox{$#1$}}\over
        \vphantom1\smash{\raise.3ex\hbox{$#2$}}}}       
\def\afrac#1#2{{\vphantom1\smash{\lower.5ex\hbox{$#1$}}\over#2}}    

\def\[{\lfloor{\hskip 0.35pt}\!\!\!\lceil}
\def\]{\rfloor{\hskip 0.35pt}\!\!\!\rceil}

\def\du#1#2{_{#1}{}^{#2}}

\def\tr{{\rm tr}}

\def\un{\underline}
\def\fracmm#1#2{{{#1}\over{#2}}}

\def\low#1{{\raise -3pt\hbox{${\hskip 0.75pt}\!_{#1}$}}}

\def\Dot#1{\buildrel{_{_{\hskip 0.01in}\bullet}}\over{#1}}
\def\dt#1{\Dot{#1}}

\def\sbar#1{\stackrel{*}{\Bar{#1}}}

\newskip\humongous \humongous=0pt plus 1000pt minus 1000pt
\def\caja{\mathsurround=0pt}
\def\eqalign#1{\,\vcenter{\openup2\jot \caja
        \ialign{\strut \hfil$\displaystyle{##}$&$
        \displaystyle{{}##}$\hfil\crcr#1\crcr}}\,}
\newif\ifdtup

\def\pl#1#2#3{Phys.~Lett.~{\bf {#1}B} (19{#2}) #3}
\def\np#1#2#3{Nucl.~Phys.~{\bf B{#1}} (19{#2}) #3}
\def\prl#1#2#3{Phys.~Rev.~Lett.~{\bf #1} (19{#2}) #3}
\def\pr#1#2#3{Phys.~Rev.~{\bf D{#1}} (19{#2}) #3}
\def\cqg#1#2#3{Class.~and Quantum Grav.~{\bf {#1}} (19{#2}) #3}

\def\ibid#1#2#3{{\it ibid.}~{\bf {#1}} (19{#2}) #3}

\parskip=\medskipamount                 
\thicklines                         

\begin{document}


\thispagestyle{empty}               

\def\border{                                            
        \setlength{\unitlength}{1mm}
        \newcount\xco
        \newcount\yco
        \xco=-24
        \yco=12
        \begin{picture}(140,0)
        \put(-20,11){\tiny Institut f\"ur Theoretische Physik Universit\"at
Hannover~~ Institut f\"ur Theoretische Physik Universit\"at Hannover~~
Institut f\"ur Theoretische Physik Hannover}
        \put(-20,-241.5){\tiny Institut f\"ur Theoretische Physik Universit\"at
Hannover~~ Institut f\"ur Theoretische Physik Universit\"at Hannover~~
Institut f\"ur Theoretische Physik Hannover}
        \end{picture}
        \par\vskip-8mm}

\def\headpic{                                           
        \indent
        \setlength{\unitlength}{.8mm}
        \thinlines
        \par
        \begin{picture}(29,16)
        \put(75,16){\line(1,0){4}}
        \put(80,16){\line(1,0){4}}
        \put(85,16){\line(1,0){4}}
        \put(92,16){\line(1,0){4}}

        \put(85,0){\line(1,0){4}}
        \put(89,8){\line(1,0){3}}
        \put(92,0){\line(1,0){4}}

        \put(85,0){\line(0,1){16}}
        \put(96,0){\line(0,1){16}}
        \put(92,16){\line(1,0){4}}

        \put(85,0){\line(1,0){4}}
        \put(89,8){\line(1,0){3}}
        \put(92,0){\line(1,0){4}}

        \put(85,0){\line(0,1){16}}
        \put(96,0){\line(0,1){16}}
        \put(79,0){\line(0,1){16}}
        \put(80,0){\line(0,1){16}}
        \put(89,0){\line(0,1){16}}
        \put(92,0){\line(0,1){16}}
        \put(79,16){\oval(8,32)[bl]}
        \put(80,16){\oval(8,32)[br]}

        \end{picture}
        \par\vskip-6.5mm
        \thicklines}

\border\headpic {\hbox to\hsize{
\vbox{\noindent DESY 97 -- 103  \hfill June 1997 \\
ITP--UH--18/97 \hfill hep-th/9706079 \\
ISSN 0418--9833 \hfill revised: May 1998 }}}

\noindent
\vskip1.3cm
\begin{center}

{\Large\bf    On the next-to-leading-order correction to 
\vglue.1in    the effective action in $N=2$ gauge theories}
\footnote{Supported in part by the `Deutsche Forschungsgemeinschaft' 
          and the NATO grant CQG 930789}\\
\vglue.3in

Sergei V. Ketov \footnote{
On leave of absence from:
High Current Electronics Institute of the Russian Academy of Sciences,
\newline ${~~~~~}$ Siberian Branch, Akademichesky~4, Tomsk 634055, Russia}

{\it Institut f\"ur Theoretische Physik, Universit\"at Hannover}\\
{\it Appelstra\ss{}e 2, 30167 Hannover, Germany}\\
{\sl ketov@itp.uni-hannover.de}
\end{center}
\vglue.2in
\begin{center}
{\Large\bf Abstract}
\end{center}

I attempt to analyse the next-to-leading-order non-holomorphic contribution
to the Wilsonian low-energy effective action in the four-dimensional $N=2$ 
gauge theories with matter, from the manifestly $N=2$ supersymmeric point of 
view, by using the harmonic superspace. The perturbative one-loop correction 
is found to be in agreement with the $N=1$ superfield calculations of de Wit, 
Grisaru and Ro\v{c}ek. The previously unknown coefficient in front of this 
non-holomorphic correction is calculated. A special attention is devoted to 
the $N=2$ {\it superconformal} gauge theories, whose one-loop non-holomorphic 
contribution is likely to be {\it exact}, even non-perturbatively. This leading 
(one-loop) non-holomorphic contribution to the LEEA of the $N=2$ superconformally 
invariant gauge field theories is calculated, and it does not vanish, similarly 
to the case of the $N=4$ super-Yang-Mills theory.

\newpage

\section{Introduction}

Extended supersymmetry severely restricts the form of the quantum effective 
action in the four-dimensional gauge theories like $N=2$ QCD. The Wilsonian
effective action to be obtained by integrating out all massive degrees of 
freedom from the fundamental (microscopic) Lagrangian is highly non-local, but
it can be expanded in powers of space-time momenta divided by the 
characteristic physical scale $\L$. Under certain physical assumptions about 
the global structure of the quantum moduli space of vacua, it becomes possible
to obtain the exact solution to the leading low-energy contribution describing
the spectrum and the {\it holomorphic} static gauge couplings in the Coulomb 
branch of the full quantum theory~\cite{sw}. The natural  step further is to 
determine the next-to-leading-order contribution to the {\it low-energy 
effective action} (LEEA), that describes {\it non-holomorphic} static 
gauge couplings. It seems to be a more complicated problem since the 
electromagneitc duality and holomorphy alone are not enough to fix 
non-holomorphic couplings. 

It is therefore desirable to understand the nature of both holomorphic and
non-holomorphic contributions to the LEEA from the fully $N=2$ supersymmetric 
point of view, when all the relevant symmetries of the microscopic Lagrangian 
are manifest. It is of particular importance to explicitly calculate quantum 
perturbative corrections by using manifestly $N=2$ supersymmetric Feynman 
rules. It would then allow one to check various proposals based on quantum 
calculations in components or $N=1$ superfields. 
The only known approach that provides an 
off-shell (model-independent) formulation for both $N=2$ gauge multiplets and 
hypermultiplets, as well as the manifestly $N=2$ supersymmetric Feynman rules 
is the $N=2$ {\it harmonic superspace} (HSS)~\cite{hs,hsf}.

The LEEA in the gauge sector of $N=2$ theories in HSS was recently studied in 
refs.~\cite{bbi,bb}, where it was shown that the one-loop holomorphic 
contributions only emerge after accounting non-vanishing central 
charges in the $N=2$ SUSY algebra. It was also demonstrated~\cite{bbi} that 
the one-loop holomorphic contribution in fact coincides with Seiberg's 
perturbative LEEA which was originally obtained by integrating the chiral 
anomaly~\cite{sei}. The same central charges also play an important role in 
the matter hypermultiplet sector of LEEA. As was shown in refs.~\cite{ke,ikz}, 
they lead to non-trivial quantum corrections to the free hypermultiplet action,
which modify the kinetic terms and generate a non-trivial scalar potential of
matter. In all these studies, HSS appears to be the indispensable tool for
 quantum calculations, since the very transparent and practical $N=2$ 
Feynman rules can be formulated only in HSS. Though the HSS method assumes the 
infinite number of ghosts and auxiliary fields in components, it does not 
really lead to complications in interpreting the results that can be reformulated in 
the conventional superspace language or in components (with a finite number of 
fields involved). The connection between the HSS approach and the more 
conventional methods is, however, non-trivial and it needs to be developed 
further. 

In sect.~2 the HSS approach is reviewed along the lines of the original 
papers~\cite{hs,hsf}. It simultaneously introduces our notation. In sect.~3
we complete the calculation of the next-to-leading-order non-holomorphic 
correction to the gauge LEEA in the $N=2$ super-QED, which was initiated in
refs.~\cite{bbi,bb}, and then generalize it to the case of the spontaneously
broken $N=2$ {\it super-Yang-Mills} (SYM) theory, including the Seiberg-Witten 
model, and add the fundamental hypermultiplet matter as well. In sect.~4, the 
results of sect.~3 are applied to the finite and scale-invariant $N=2$ 
supersymmetric gauge theories. The relevance of the perturbative non-holomorphic 
contributions to the LEEA of finite $N=2$ supersymmetric gauge field theories in 
four dimensions for checking M-theory was recently emphasized by Dine and 
Seiberg in ref.~\cite{ds}.
\vglue.2in

\section{Basic facts about $N=2$ gauge theories in HSS}

In the HSS formalism, the standard N=2 superspace 
$z=(x^m,\q^{\a}_i,\bar{\q}^{\dt{\a}i})$, where $m=0,1,2,3$, $\a=1,2$, 
and $i=1,2$,
is extended by adding the bosonic variables (or `zweibeins') $u^{\pm i}$ 
parametrizing the sphere $S^2\sim SU(2)/U(1)$. The $SU(2)$ indices are raised 
and lowered with the antisymmetric Levi-Civita symbols $\ve_{ij}$ and  
$\ve^{ij}$, $\ve^{12}=-\ve_{12}=1$. The ordinary complex conjugation 
is detoned by bar. One has
$$ \left( \begin{array}{c} u^{+i} \\ u^{-i}\end{array}\right) \in SU(2)~,
\quad {\rm so~~that}\quad u^{+i}u^-_i=1~,\quad{\rm and}\quad
 u^{+i}u^+_i=u^{-i}u^-_i=0~.\eqno(2.1)$$

Instead of employing an explicit parametrization of the sphere, 
it is convenient 
to deal with functions of zweibeins, that carry a definite $U(1)$ charge $q$ 
to be defined by $q(u^{\pm}_i)=\pm 1$, and use the following integration 
rules~\cite{hs}:
$$ \int du =1~,\qquad \int du\, u^{+(i_1}\cdots u^{+i_m}u^{-j_1}\cdots
u^{-j_n)}=0~,\quad {\rm when}\quad m+n>0~.\eqno(2.2)$$
It is obvious that the integral over a $U(1)$-charged quantity vanishes.  

In addition to the usual complex conjugation, there exists a star conjugation 
that only acts on the $U(1)$ indices, $(u^+_i)^*=u^-_i$ and
$(u^-_i)^*=-u^+_i$. One easily finds~\cite{hs}
$$ \sbar{u^{\pm i}}=-u^{\pm}_i~,\qquad  \sbar{u^{\pm}_i}=u^{\pm i}~.
\eqno(2.3)$$

The covariant derivatives with respect to the zweibeins,  that preserve the 
defining conditions (2.1), are given by
$$ D^{++}=u^{+i}\fracmm{\pa}{\pa u^{-i}}~,\quad 
D^{--}=u^{-i}\fracmm{\pa}{\pa u^{+i}}~,\quad 
D^{0}=u^{+i}\fracmm{\pa}{\pa u^{+i}}-u^{-i}\fracmm{\pa}{\pa u^{-i}}~.
\eqno(2.4)$$
It is easy to check that they satisfy the $SU(2)$ algebra,
$$\[ D^{++},D^{--}\]=D^0~,\quad \[D^0,D^{\pm\pm}\]=\pm 2D^{\pm\pm}~.
\eqno(2.5)$$

The key feature of the $N=2$ HSS is the existence of the {\it analytic}
subspace parametrized by the coordinates
$$ (\z,u)=\left\{ \begin{array}{c}
x^m_{\rm A}=x^m-2i\q^{(i}\s^m\bar{\q}^{j)}u^+_iu^-_j~,~~
\q^+_{\a}=\q^i_{\a}u^+_i~,~~ \bar{\q}^+_{\dt{\a}}=\bar{\q}^i_{\dt{\a}}u^+_i~,~~
u^{\pm}_i \end{array} \right\}~,\eqno(2.6)$$
which is invariant under $N=2$ supersymmetry, and is closed under the combined 
conjugation of eq.~(2.3)~\cite{hs}. It allows one to define the {\it analytic} 
superfields of any $U(1)$ charge $q$, by the analyticity 
conditions
$$D^+_{\a}\f^{(q)}=\bar{D}^+_{\dt{\a}}\f^{(q)}=0~,\quad {\rm where}\quad
D^+_{\a}=D^i_{\a}u^+_i \quad {\rm and}\quad
\bar{D}^+_{\dt{\a}}=\bar{D}^i_{\dt{\a}}u^+_i~,\eqno(2.7)$$
and introduce the analytic measure $d\z^{(-4)}du\equiv d^4x_{\rm A}
d^2\q^+d^2\bar{\q}^+du$ of charge $(-4)$, so that the full measure in 
the $N=2$ HSS can be written down as
$$ d^4xd^4\q d^4\bar{\q}du=d\z^{(-4)}du(D^+)^4~,\eqno(2.8)$$
where
$$(D^+)^4=\fracmm{1}{16}(D^+)^2(\bar{D}^+)^2
=\fracmm{1}{16}(D^{+\a}D_{\a}^+)(\bar{D}^{+}_{\dt{\a}}\bar{D}^{+\dt{\a}})~.
\eqno(2.9)$$
In the analytic subspace, the harmonic derivative 
$$D^{++}_{\rm A}=D^{++}-2i\q^+\s^m\bar{\q}^+\pa_m\eqno(2.10)$$ 
preserves analyticity and allows one to integrate by parts. Since both the
original (central) basis and the analytic one can be used on equal footing 
in HSS, in what follows we omit the subscript A at the covariant derivatives
in the analytic basis.

It is the advantage of the analytic HSS that both a (massless) hypermultiplet 
and an $N=2$ vector multiplet can be introduced there on equal footing. 
Namely, the hypermultiplet can be defined as an unconstrained complex analytic
superfield $q^+$ of the $U(1)$-charge $(+1)$, whereas the $N=2$ vector 
multiplet is described by an unconstrained analytic superfield $V^{++}$ of the
$U(1)$-charge $(+2)$. The $V^{++}$ is real in the sense 
$\Bar{V^{++}}^{\,*}=V^{++}$, and it can be naturally introduced 
as a connection to the harmonic derivative $D^{++}$. 
The both fields, $q^+$ and $V^{++}$, can be Lie 
algebra-valued in a fundamental or adjoint representation of the gauge group. 
The hypermultiplet action with a minimal coupling to the gauge superfield reads
$$ S[q,V]= -\tr\int d\z^{(-4)}du \,\sbar{q}{}^+(D^{++}+iV^{++})q^+~.
\eqno(2.11)$$
It is not difficult to check that the free hypermultiplet equations of motion, 
$D^{++}q^+=0$ imply $q^+=q^i(z)u^+_i$ and the usual (on-shell)
Fayet-Sohnius constraints~\cite{fs} in the ordinary $N=2$ superspace,
$$D_{\a}^{(i}q^{j)}(z)=D_{\dt{\a}}^{(i}q^{j)}(z)=0~.\eqno(2.12)$$

There exists another, equivalent HSS description of a massless hypermultiplet 
in terms of an unconstrained analytic superfield $\o$ with the vanishing 
$U(1)$-charge~\cite{hs}, and the action
$$S[\o,V]=-\frac{1}{2}\tr\int d\z^{(-4)}du \,
(D^{++}+iV^{++})\o(D^{++}+iV^{++})\o~.\eqno(2.13)$$
The corresponding ($V$-dependent) effective actions to be obtained by 
integrating over the hypermultiplet $q^+$ or $\o$, respectively, are the same 
when one trades each $q$ hypermultiplet for a real $\o$ 
hypermultiplet~\cite{bbi}. The off-shell HSS hypermultiplet in terms of the 
ordinary $N=2$ superfields is just the infinitely-relaxed $N=2$ tensor 
multiplet~\cite{krev}.  

The $N=2$ SYM theory is usually formulated in the ordinary $N=2$ superspace by
imposing certain constraints on the gauge- and super- covariant derivatives
${\cal D}^i_{\a}$ and $\bar{\cal D}^i_{\dt{\a}}$ ~\cite{gsw}. The constraints 
\cite{gsw}, in essence, boil down to the existence of a covariantly chiral and
gauge-covariant $N=2$ SYM field strength $W$ satisfying the 
reality condition (Bianchi `identity') 
$$ {\cal D}^{\a}\low{(i}{\cal D}\low{j)\a}W=\bar{\cal D}_{\dt{\a}(i}
\bar{\cal D}^{\dt{\a}}\low{j)}\bar{W}~.\eqno(2.14)$$

Unlike the $N=1$ SYM theory, an $N=2$ supersymmetric solution to the 
non-abelian
 $N=2$ SYM constraints in the ordinary $N=2$ superspace is not known in an 
analytic form. It is the $N=2$ HSS reformulation of the $N=2$ SYM theory that
makes it possible~\cite{hs}. The exact non-abelian relation between the 
constrained, harmonic-independent superfield strength $W$ and the unconstrained
analytic superfield $V^{++}$ is given in refs.~\cite{hs,hsf}, and it is highly
non-linear. It is merely its abelian version that is needed for calculating 
the perturbative LEEA. It is not difficult to check, or just use the results 
of ref.~\cite{zupnik}, that the abelian relation takes the form
$$ W=\fracmm{1}{4} \{ \bar{\cal D}^+_{\dt{\a}},\bar{\cal D}^{-\dt{\a}}\}
=-\fracmm{1}{4}(\bar{D}^+)^2A^{--}~,\eqno(2.15)$$
where the non-analytic harmonic superfield connection $A^{--}(z,u)$ to the 
derivative $D^{--}$ has been introduced, ${\cal D}^{--}=D^{--}+iA^{--}$.
As a consequence of the $N=2$ HSS abelian constraint 
$\[{\cal D}^{++},{\cal D}^{--}\]={\cal D}^0=D^0$, the connection $A^{--}$ 
satisfies the relation
$$ D^{++}A^{--}=D^{--}V^{++}~,\eqno(2.16)$$
whereas eq.~(2.14) can be rewritten to the form
$$ (D^+)^2W=(\bar{D}^+)^2\bar{W}~.\eqno(2.17)$$

A solution to the $A^{--}$ in terms of the analytic unconstrained superfield
$V^{++}$ easily follows from eq.~(2.16) when using the identity~\cite{hsf}
$$ D^{++}_1(u_1^+u^+_2)^{-2}=D_1^{--}\d^{(2,-2)}(u_1,u_2)~,\eqno(2.18)$$
where we have introduced the harmonic delta-function $\d^{(2,-2)}(u_1,u_2)$ 
and the harmonic distribution $(u_1^+u^+_2)^{-2}$ according to their 
definitions in refs.~\cite{hs,hsf}, 
hopefully, in the self-explaining notation. One finds~\cite{zupnik} 
$$ A^{--}(z,u)= \int dv \,\fracmm{V^{++}(z,v)}{(u^+v^+)^2}~,\eqno(2.19)$$
and
$$ W(z)=-\fracmm{1}{4}\int du (\bar{D}^-)^2V^{++}(z,u)~,\quad \bar{W}(z)=
-\fracmm{1}{4}\int du (D^-)^2V^{++}(z,u)~,\eqno(2.20)$$
by using the identity 
$$u^+_i=v^+_i(v^-u^+)-v^-_i(u^+v^+)~,\eqno(2.21)$$
which is the obvious consequence of the definitions (2.1). 

The equations of motion are given by the vanishing analytic superfield
$$ (D^+)^4A^{--}(z,u)=0~,\eqno(2.22)$$
while the corresponding action reads~\cite{zupnik}
$$ \eqalign{
S[V]= & \fracmm{1}{4}\int d^4xd^4\theta\, W^2 + {\rm h.c.}=
\fracmm{1}{2}\int d^4xd^4\theta d^4\bar{\theta}du \,V^{++}(z,u)A^{--}(z,u)\cr
= & \fracmm{1}{2}\int d^4xd^4\theta d^4\bar{\theta}du_1du_2\,
\fracmm{V^{++}(z,u_1)V^{++}(z,u_2)}{(u_1^+u^+_2)^2}~.\cr}\eqno(2.23)$$

In a WZ-like gauge, the abelian analytic pre-potential $V^{++}$ amounts 
to~\cite{hs}
$$\eqalign{
 V^{++}(x_{\rm A},\theta^+,\bar{\theta}^+,u)=&
\bar{\theta}^+\bar{\theta}^+a(x_{\rm A})
+ \bar{a}(x_{\rm A})\theta^+\theta^+ 
-2i\theta^+\s^{m}\bar{\theta}^+V_{m}(x_{\rm A}) \cr
& +\bar{\theta}^+\bar{\theta}^+\theta^{\a +}\j^i_{\a}(x_{\rm A})u^-_i
+\theta^+\theta^+\bar{\theta}^+_{\dt{\a}}\bar{\j}^{\dt{\a}i}(x_{\rm A})u^-_i\cr
&+\theta^+\theta^+\bar{\theta}^+\bar{\theta}^+D^{(ij)}(x_{\rm A})u^-_iu^-_j~,
\cr} \eqno(2.24)$$
where $(a,\j^i_{\a},V_{m},D^{ij})$ are the usual $N=2$ vector multiplet 
components~\cite{gsw}.

The (BPS) mass of a hypermultiplet can only come from the central charges 
of the 
$N=2$ SUSY algebra since, otherwise, the number of the massive hypermultiplet 
components has to be increased. The most natural way to introduce central 
charges $(Z,\bar{Z})$ is to identify them with spontaneously broken $U(1)$ 
generators of dimensional reduction from six dimensions via the Scherk-Schwarz
mechanism~\cite{ss}. Being rewritten to six dimensions, 
eq.~(2.10) implies the additional `connection' term in the associated 
four-dimensional harmonic derivative 
$$ {\cal D}^{++}=D^{++}+v^{++}~,\quad {\rm where}\quad 
v^{++}=i(\theta^+\theta^+)\bar{Z}+i(\bar{\theta}^+\bar{\theta}^+)Z~.
\eqno(2.25)$$
Comparing eq.~(2.25) with eqs.~(2.11) and (2.20) clearly shows that the $N=2$ 
central charges can be equivalently treated as a non-trivial $N=2$ gauge
background, with the covariantly constant chiral superfield strength
$$ \VEV{W}=\VEV{a}=Z~,\eqno(2.26)$$
where eq.~(2.24) has been used too.
\vglue.2in

\section{The one-loop LEEA for $N=2$ gauge fields}

The gauge-invariant hypermultiplet action in the $N=2$ SYM background with 
the gauge group $SU(2)$ is to be supplemented by a 
gauge-fixing term and the corresponding ghost terms. The $N=2$ ghost structure 
within the background-field
method in HSS was recently studied in ref.~\cite{bb}. As was shown in 
ref.~\cite{bb} there are two types of ghosts in the adjoint representation of 
$SU(2)$: the Faddeev-Popov (FP) fermionic ghosts to be represented by two real 
$\o\low{\rm FP}$-hypermultiplets, and the Nielsen-Kallosh (NK) bosonic ghosts 
to be represented by a real $\o\low{\rm NK}$-hypermultiplet. 
Most importantly, the $N=2$ SYM one-loop effective action in HSS is found 
to be entirely determined by the ghost contributions alone~\cite{bb}. 

In the Coulomb branch of the quantum theory, the gauge group $SU(2)$ is broken
to its $U(1)$ subgroup so that only an abelian $N=2$ gauge component 
represents the light degrees of freedom at a generic point in the moduli 
space of vacua,~\footnote{We use the normalization condition 
$\tr(t^at^b)=\d^{ab}$ for the $SU(2)$ generators $t^a$, $a=1,2,3$, in 
\newline ${~~~~~}$ {\it any} representation. The $SU(2)$ gauge coupling constant
is set to be $e^2=2$.} 
$$ V^{++}\equiv \fracmm{1}{\sqrt{2}}\t^a V^{a++}\to 
\fracmm{1}{\sqrt{2}} \t^3 V^{3++}~,\eqno(3.1)$$
where Pauli matrices $\t^a$ satisfy the relations 
$\[\t^a,\t^b\]=2i\ve^{abc}\t^c$ and $\tr(\t^a\t^b)=2\d^{ab}$. 
When being only interested in calculating the leading perturbative correction 
to the effective action, we do not have to integrate over the massive gauge 
fields, so that they can be simply dropped out of the microscopic Lagrangian. 
A real hypermultiplet $\o$ of unit charge, in the adjoint of $SU(2)$ then 
yields a {\it complex} hypermultiplet of charge $\sqrt{2}$, minimally coupled 
to the $U(1)$ gauge field $V^{3++}$, since one of the hypermultiplet 
components $(\o^3)$ decouples. Therefore, the total purely $N=2$ 
SYM contribution to the LEEA is given by {\it minus} that of a real bosonic 
$\o$-hypermultiplet in the adjoint or, equivalently, by {\it minus twice} that 
of a single bosonic $q$-hypermultiplet of charge $\sqrt{2}$~\cite{bb}. 
Similarly, the LEEA contribution of a matter $q$-hypermultiplet with unit 
charge, in the fundamental representation of $SU(2)$ is {\it twice} that of a 
single $q$-hypermultiplet of charge $1/\sqrt{2}$.

Having reduced a calculation of the perturbative LEEA in the gauge sector to 
the {\it abelian} problem for a single $q$-hypermultiplet minimally coupled 
to the background $U(1)$ gauge superfield $V^{++}$ in HSS, we now have to 
integrate over the $q$-hypermultiplet. The corresponding basic effective 
action $\G_q[V]$ is given by a sum of the one-loop HSS graphs in powers of the
background gauge superfield, and the loop to be constructed out of the 
hypermultiplet propagators (Fig.~1). 

\begin{figure}[t]
\vglue.1in
\makebox{
\epsfxsize=4in
\epsfbox{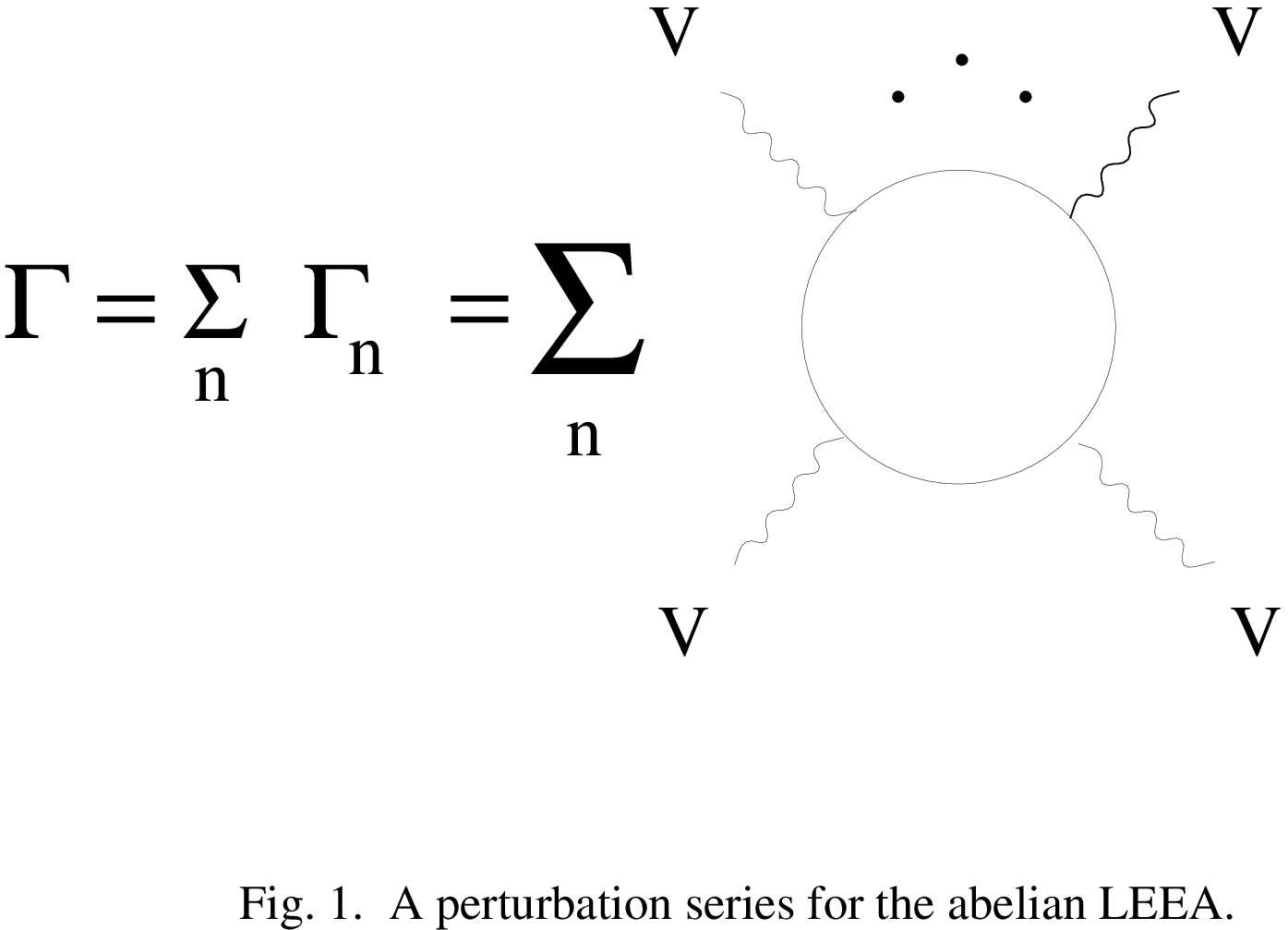}
}
\end{figure}

It is straightforward to calculate each HSS graph, by using either the massive
$q$-hypermultiplet propagator with the BPS mass~\cite{ikz},
$$ i\VEV{q^+(1)\sbar{q}{}^+(2)}=-\,\fracmm{1}{\bo_1+Z\bar{Z}}(D_1^+)^4(D_2^+)^4
e^{v_2-v_1}\d^{12}(z_1-z_2)\fracmm{1}{(u^+_1u_2^+)^3}~,\eqno(3.2a)$$
where 
$$ v\equiv i(\q^+\q^-)\bar{Z}+i(\bar{\q}^+\bar{\q}^-)Z~,\eqno(3.2b)$$
or, equivalently,
the massless one $(Z=v=0)$ but with the special background gauge superfield, 
$iV^{++}\to iV^{++}+v^{++}$, where $v^{++}$ is given by eq.~(2.25). 

The calculation goes as follows~\cite{bbi}. First, one restores the full 
Grassmann measure by taking the factors $(D^+_1)^4\cdots (D^+_n)^4$ off the 
hypermultiplet propagators. It allows one to explicitly integrate over all but
one set of the anticommuting HSS coordinates by using the Grassmann 
delta-functions, thus obtaining the full Grassmann measure $d^8\theta$. 
Further integrating by parts and using eq.~(2.21), one can cancel the harmonic
distribution $(u^+_1u^+_2)^{-3}(u^+_2u^+_3)^{-3}\cdots (u^+_{n-1}u^+_n)^{-3}
(u^+_nu^+_1)^{-3}$, which simultaneously means the absence of potential 
divergences in the harmonic variables, in accordance with the general analysis
of ref.~\cite{hsf}. Because of the gauge invariance, the remaining local terms
in the low-energy approximation can only depend upon $W$ and $\bar{W}$ via 
eq.~(2.20). It allows one to eliminate all the dependence upon the harmonic 
variables in the LEEA, and end up with
$$\G[V]=\left[ \int d^4x d^4\theta \,\cf(W) +{\rm h.c.}\right]
+ \int d^4x d^4\theta d^4\bar{\theta}\,\ch(W,\bar{W})~.\eqno(3.3)$$

The holomorphic contribution to the one-loop LEEA appears due to the 
non-vanishing central charges whose presence gives rise to the 
$(\theta^-)^4$-dependent terms before Grassmann integration in the HSS graphs.
As was shown in 
ref.~\cite{bbi}, these terms deliver Seiberg's perturbative LEEA~\cite{sei},
$$ \cf_q(W)=-\,\fracmm{1}{32\p^2}W^2\ln\fracmm{W^2}{M^2}~,\eqno(3.4)$$
where the renormalization scale $M$ is fixed by the condition $\cf_q(M)=0$. 
Note that the result (3.4) does not depend upon an infra-red cutoff $\L$. 

Similarly, the non-holomorphic perturbative contribution from the one-loop HSS 
graphs is given by~\footnote{The infra-red cutoff $\L$ is fixed by
the condition $\ch_q(\L,\L)=0$.}
$$ \ch_q(W,\Bar{W})=\fracmm{1}{(16\p)^2} \sum^{\infty}_{k=1}\fracmm{(-1)^{k+1}}{k^2}\left(
\fracmm{W\Bar{W}}{\L^2}\right)^k=\fracmm{1}{(16\p)^2}\int_0^{W\Bar{W}/\L^2}\,\fracmm{d\x}{\x}
\ln(1+\x)~,\eqno(3.5)$$
where we have used the standard integral representation for the dilogarithm function. The first
term in eq.~(3.5) was calculated in ref.~\cite{bbi}, where it was interpreted as the
$N=2$ supersymmetric Heisenberg-Euler Lagrangian.

It is not difficult to verify that the asymptotical perturbation series (3.5) can be rewritten to
the form suggested for the abelian case by de Wit, Grisaru and Ro\v{c}ek in ref.~\cite{wgr},
$$ \ch_q(W,\Bar{W})=\fracmm{1}{(16\p)^2}\ln\left(\fracmm{W}{\L}\right)
\ln\left(\fracmm{\Bar{W}}{\L}\right)~,\eqno(3.6)$$
or, equivalently, as (see eq.~(4.4) below)
$$\ch_q(W,\Bar{W})=\fracmm{1}{2(16\p)^2}\ln^2\left(\fracmm{W\Bar{W}}{\L^2}\right)~,\eqno(3.7)$$
where we have used the fact that the non-holomorphic function $\ch$ is defined modulo the K\"ahler 
gauge transformations
$$\ch(W,\Bar{W})\to \ch(W,\Bar{W})+f(W) +\bar{f}(\Bar{W})~,
\eqno(3.8)$$
with an arbitrary holomorphic function $f(W)$ as a parameter. The non-holomorphic contribution of
eq.~(3.6) or (3.7) does not really depend upon the scale $\L$, again due to the K\"ahler
invariance (3.8).

The HSS result (3.4) for the perturbative part of the holomorphic (SW) $N=2$ gauge LEEA agrees with
the Seiberg argument \cite{sei} based on the perturbative $U(1)_R$ symmetry and an integration of 
the associated chiral anomaly. As is obvious from eq.~(3.6), the next-to-leading-order 
non-holomorphic contribution to the SW gauge LEEA satisfies a simple differential equation
$$ W\Bar{W}\pa_W\pa_{\Bar{W}}\ch_q(W,\Bar{W})=const.~,\eqno(3.9)$$
which can be considered as the direct consequence of scale and $U(1)_R$ invariances~\cite{ds}.

Since the Seiberg result~\cite{sei} for the perturbative holomorphic function $\cf(W)$ was 
indirectly obtained by integrating the chiral anomaly, its direct and manifestly $N=2$ 
supersymmetric HSS derivation \cite{bbi} means, in particular, that the one-loop contribution 
(3.4) {\it exactly} saturates the corresponding anomalous perturbative LEEA. 

It is natural to replace $\L$ in eq.~(3.6) with a field-dependent cutoff $a$ whose vacuum
expectation value $\VEV{a}$ parameterizes the moduli space of vacua, and can be identified 
with the central charge because of eq.~(2.26),
$$ \ch_q(W,\bar{W})=\fracmm{1}{2(16\p)^2}
\ln^2\left(\fracmm{W\bar{W}}{a\bar{a}}\right)~.\eqno(3.10)$$

As a simple application, consider the celebrated Seiberg-Witten model,
whose fundamental (microscopic) action describes the purely gauge $N=2$ SYM 
theory, with the $SU(2)$ gauge group spontaneously broken to its $U(1)$
subgroup~\cite{sw}. The perturbative low-energy effective action reads
$$\eqalign{
\G^{\rm SW}_{\rm perturbative} &=
-4\int_{\rm chiral}\,\cf_q(W) -2\int \ch_q(W,\bar{W}) \cr
&= +\fracmm{1}{(4\p)^2}\int_{\rm chiral}\,W^2\ln\fracmm{W^2}{M^2}
-\fracmm{1}{(16\p)^2}\int \ln^2\left(\fracmm{W\bar{W}}{a\bar{a}}\right)~.
\cr}\eqno(3.11)$$

In the strong coupling region, near a singularity in the quantum 
moduli space where a BPS-like (t'Hooft-Polyakov) monopole becomes massless, 
the Seiberg-Witten model takes the form of the {\it dual} $N=2$ supersymmetric
QED after the duality transformation $V^{++}\rightarrow V^{++}_D$, 
$W\rightarrow W_D$ and $a\to a_D$. The t'Hooft-Polyakov monopole is known to 
belong to a $q^+$-hypermultiplet 
that represents the non-perturbative degrees of freedom in the 
theory~\cite{sw}. Therefore, the low-energy effective action near the 
monopole singularity is given by
$$\eqalign{
\G^{\rm SW}_D & =+\int_{\rm chiral}\,\cf_q(W_D) +\int \ch_q(W_D,\bar{W}_D)\cr
& = -\fracmm{1}{32\p^2}\int_{\rm chiral}\,W_D^2\ln\fracmm{W^2_D}{M^2_D}
+\fracmm{1}{2(16\p)^2}\int \ln^2\left(\fracmm{W_D\bar{W}_D}{a_D\bar{a}_D}
\right)~.\cr}\eqno(3.12)$$

The exact holomorphic low-energy effective action~\cite{sw} is known to have 
just two physical singularities in the quantum moduli space, where a BPS-like 
particle becomes massless. The perturbative next-to-leading-order term 
$\ch(W,\bar{W})$ is not singular at that points. Of course, we still have to
justify its use at strong coupling (see the next sect.~4). The 
$SL(2,{\bf Z})$ duality requires the {\it exact} function $\ch(W,\bar{W})$ to 
be duality-invariant~\cite{henn}. Being combined with the perturbative
information from eqs.~(3.11) and (3.12), it is, however, not enough to 
determine the exact form of that function (see ref.~\cite{matone} for some
additional proposals).

In a more general case of $N_f$, $q^+$-type hypermultiplets in the fundamental representation of the
gauge group $SU(N_c)$, i.e. the $N=2$ super-QCD, the extra coefficient in front of the holomorphic
contribution $\cf$ is proportional to the one-loop RG beta-function $(N_f-2N_c)$, whereas the
extra coefficient in front of the non-holomorphic contribution $\ch$ is proportional to
$(2N_f-N_c)$, in the $N=2$ super-Feynman gauge used above. In another interesting case of the $N=4$
super-Yang-Mills theory, whose $N=2$ matter content is given by a single $\o$-type
hypermultiplet in the {\it adjoint} representation of the gauge group, the numerical coefficient in
front of the holomorphic function $\cf$ vanishes together with the RG beta-function, whereas the
numerical coeffient in front of the non-holomorphic contribution $\ch$ always appears to be
positive, in agreement with the earlier calculations in terms of $N=1$ superfields (see page 390 of
ref.~\cite{super}) and some recent $N=2$ supersymmetric calculations by different 
methods~\cite{stony,bkuz}. In particular, in the case of finite and $N=2$
superconformally invariant gauge field theories $(N_f=2N_c)$, the leading non-holomorphic
contribution to the LEEA is given by eq.~(3.5) multiplied by $3N_c$, and it never vanishes.
\vglue.2in

\section{Instanton corrections}

The exact solution to the holomorphic LEEA was obtained by Seiberg and Witten
~\cite{sw} by the use of duality and holomorphicity properties of the $N=2$ 
SYM theory with the $SU(2)$ gauge group, under certain physical assumptions 
about the global structure of the quantum moduli space of vacua. The
Seiberg-Witten solution is encoded in terms of the auxiliary elliptic curve
$$ y^2=(x^2-u)^2-\L_{\rm SW}^4~,\eqno(4.1)$$
where the moduli space parameter $u$ can be identified with the expectation
value of the gauge-invariant operator, $u=\VEV{\tr\,a^2}$, and $\L_{\rm SW}$ is
the renormalisation group invariant (Seiberg-Witten) scale. The holomorphic 
function $\cf(W)$ can then be parametrized in terms of the periods of certain
abelian differentials of the 3rd kind, associated with the SW curve (4.1),
see e.g., ref.~\cite{ketov} for a review. When being expanded in the inverse 
powers of $W$ at weak coupling (near $u=\infty$), 
the Seiberg-Witten solution reads
$$ \cf_{\rm per.}(W)=\fracmm{1}{32\p^2}\left[
W^2\ln\fracmm{W^2\sqrt{2}}{\L_{\rm SW}^2}-\fracmm{(\L_{\rm SW}/2)^4}{W^2}
+\ldots\right]~,\eqno(4.2)$$
where the first term coincides with that in eq.~(3.11) after identifying
$M^2=\L_{\rm SW}^2/\sqrt{2}$, whereas the rest of terms represents 
multi-instanton corrections
(the leading one-instanton correction is explicitly written down). The latter 
can be computed independently~\cite{fp}, in agreement with the exact 
Seiberg-Witten result. The large-distance instanton effects can be encoded in 
terms of the effective {\it Callan-Dashen-Gross} (CDG) vertex to be added to 
the microscopic Lagrangian~\cite{cdg}. Its $N=2$ supersymmetric generalization
 was constructed by Yung~\cite{yung1}, who also applied it to explicitly 
calculate the one-instanton corrections to the effective action from the first 
principles. When being restricted to 
the light degrees of freedom which are relevant for the LEEA, the momentum 
expansion of the Yung vertex confirms eq.~(4.2) and also yields the 
one-instanton correction (in the Pauli-Villars regularization scheme) to the 
non-holomorphic LEEA of the Seiberg-Witten model~\cite{yung1},
$$ \ch(W,\bar{W})_{\rm per.+one~inst.}=\fracmm{1}{(8\p)^2}\left(
-\abs{\ln\fracmm{W}{\L_{\rm SW}}}^2 
+ \fracmm{\L_{\rm SW}^4}{2W^4}\ln\fracmm{\bar{W}}{\L_{\rm SW}} 
+ \fracmm{\L_{\rm SW}^4}{2\bar{W}^4}\ln\fracmm{W}{\L_{\rm SW}}+\ldots\right)~,
\eqno(4.3)$$
where the dots stand for higher multi-instanton corrections. In rewriting
eq.~(4.3) we have used the identity
$$ \ln^2\left(\fracmm{W\bar{W}}{\L^2}\right)=2\ln\fracmm{W}{\L}
\ln\fracmm{\bar{W}}{\L} +\ln^2\fracmm{W}{\L}+\ln^2\fracmm{\bar{W}}{\L}~,
\eqno(4.4)$$
and kept only the non-holomorphic terms because of eq.~(3.8). The relative 
easyness of getting the 
perturbative and one-instanton corrections to the LEEA versus the 
multi-instanton ones is related to the fact that the former can be calculated 
by dropping the heavy gauge fields in the microscopic (CDG-modified) 
Lagrangian.
 
Being unable to calculate the exact function $\ch(W,\bar{W})$ in the 
Seiberg-Witten model at strong coupling, we can, nevertheless, ask whether at 
certain circumstances the one-loop perturbative results could be exact. In the
case with the $N=2$ matter to be represented by $N_f$ hypermultiplets in the 
fundamental representation of $SU(2)$, the corresponding elliptic curve 
generalizing that of eq.~(4.1) is given by~\cite{ho,aps}~\footnote{When 
$2<N_f\leq 4$, there are some ambiguities in the form of elliptic 
curve~\cite{ho,aps}.}
$$ y^2=(x^2-u)^2-\L_{\rm SW}^{4-N_f}\prod_{j=1}^{N_f}(x-m_j)~,\eqno(4.5)$$
where $m_j$ are hypermultiplet masses. The dependence upon $\L$ disappears at 
$N_f=4$, which corresponds to the finite and scale-invariant $N=2$~ gauge 
theories. In this case, there can be neither higher-loop perturbative 
corrections since they are dependent upon the normalization scale,~\footnote{
The absence of higher-loop perturbative corrections to the holomorphic LEEA
was verified at \newline ${~~~~~}$ two loops in ref.~\cite{ggruz}.} 
nor non-perturbative instanton contributions since they are all proportional 
to the positive powers of $\L$ (see also ref.~\cite{ds}, as well as some 
more checks in ref.~\cite{wales}). Hence, the one-loop non-holomorphic 
contribution to the LEEA is exact when $N_f=4$. In accordance with the results
of sect.~3, it is given by
$$ \ch_{\rm finite}(W,\bar{W})
=\fracmm{3}{256\p^2}\ln^2\left(\fracmm{W\bar{W}}{\VEV{W}\VEV{\bar{W}}}
\right)~.\eqno(4.6)$$
Note that eq.~(4.6) does not depend upon any scale.

Another interesting limit, where the perturbative results of sect.~3 may be 
exact is to take $\L\to 0$ at $1\leq N_f\leq 4$. 
By tuning the bare hypermultiplet
masses, one can arrange the situation when the (singular) points in the 
Coulomb branch, where some of the non-perturbative states (like monopoles or 
dyons) become massless, coincide at the so-called Argyres-Douglas 
point~\cite{adu}. It should lead to a new physics since these BPS-like 
physical states are mutually {\it non-local}. The latter means that there is no
duality transformation to another field description of these states where the
corresponding fields would have no magnetic charges.  As was argued in 
ref.~\cite{aw}, it yields new non-trivial $N=2$ superconformally invariant 
gauge field theories. To control the theory at a non-trivial Argyres-Douglas 
fixed point, it was important for the analysis of ref.~\cite{aw} to have a 
path to the Higgs branch that touches the Coulomb branch at the phase 
transition points where some of the hypermultiplets become massless. There is 
no Higgs branch at $N_f=0$ or $1$, but is appears at $N_f=2$ or $3$. It may 
not be accidental that the coefficient $(N_f-1)$ in front of the non-holomorphic 
term is related to it. The well-known fact that the one-instanton 
holomorphic correction vanishes when there is a massless matter 
hypermultiplet, may also be related to the Higgs branch in the full quantum 
theory. 
\vglue.2in


\begin{thebibliography}{99}

\bibitem{sw} N. Seiberg and E. Witten, \np{426}{94}{19}, \ibid{431}{94}{484}.
\bibitem{hs} A. Galperin, E. Ivanov, S. Kalitzin, V. Ogievetsky and
               E. Sokatchev, \cqg{1}{84}{469}.
\bibitem{hsf} A. Galperin, E. Ivanov, V. Ogievetsky and
              E. Sokatchev. \cqg{2}{85}{601}, \ibid{2}{85}{617}.
\bibitem{bbi}  I. Buchbinder, E. Buchbinder, E. Ivanov, S. Kuzenko and
B. Ovrut, \newline \pl{412}{97}{309}.
\bibitem{bb} I. Buchbinder, E. Buchbinder, S. Kuzenko and B. Ovrut,\newline \pl{417}{98}{61}.
\bibitem{sei} N. Seiberg, \pl{206}{88}{75}.
\bibitem{ke} S. Ketov, \pl{399}{97}{91}.
\bibitem{ikz} E. Ivanov, S. Ketov and B. Zupnik, \np{509}{98}{53}.
\bibitem{ds} M. Dine and N. Seiberg, \np{458}{96}{445}; \pl{409}{97}{239}.
\bibitem{fs} P. Fayet, \np{113}{76}{135};\\
             M. Sohnius, \np{138}{78}{109}.
\bibitem{krev} S. Ketov, Fortschr. Phys. {\bf 36} (1988) 361.
\bibitem{gsw} R. Grimm, M. Sohnius and J. Wess, \np{133}{78}{275}.
\bibitem{zupnik} B. Zupnik, Teor. Mat. Fiz. {\bf 69} (1986) 207. 
\bibitem{ss} J. Scherk and J. Schwarz, \np{153}{79}{61}.
\bibitem{wgr} B. de Wit, M. Grisaru and M. Ro\v{c}ek, \pl{374}{96}{297}.
\bibitem{henn} M. Henningson, \np{458}{96}{445}.
\bibitem{matone} M. Matone, \prl{78}{97}{1412}.
\bibitem{super} S. J. Gates Jr., M. T. Grisaru, M. Ro\v{c}ek and W. Siegel, {\it Superspace},
Benjamin/Cummings Publ. Comp., 1983, p.~390.
\bibitem{stony} F. Gonzalez-Rey and M. Ro\v{c}ek, {\it Non-holomorphic N=2 terms in N=4
 SYM: 1-loop calculation in N=2 superspace}, Stony Brook preprint ITP-SB-98-25; hep-th/9804010.
\bibitem{bkuz} I. Buchbinder and S. Kuzenko, {\it Comments on the background field method
in harmonic superspace: non-holomorphic corrections in N=4 SYM}, Tomsk preprint TP-TPSU-2/98 and
TP-TSU-7/98; hep-th/9804168.
\bibitem{ketov} S. Ketov, Fortschr. Phys. {\bf 45} (1997) 237.
\bibitem{fp} D. Finnell and P. Pouliot, \np{453}{95}{225};\\
N. Dorey, V. Khoze and M. Mattis, \pr{54}{96}{2941}; \\
\pl{388}{96}{324};\\
F. Fucito and G. Travaglini, \pr{55}{97}{1099}.
\bibitem{cdg} C. Callan, R. Dashen and D. Gross, \pr{17}{78}{2717};
\pr{19}{79}{1826}.
\bibitem{yung1} A. Yung, \np{485}{97}{38}.
\bibitem{ho} A. Hanany and Y. Oz, \np{452}{95}{283}.
\bibitem{aps} P. Argyres, M. Plesser and A. Shapere, \prl{75}{95}{1699}.
\bibitem{ggruz} A. De Giovanni, M. Grisaru, M. Ro\v{c}ek, R. von Unge and
D. Zanon,\newline \pl{409}{97}{251}.
\bibitem{wales} N. Dorey, V. Khoze, M. Mattis, M. Slater and W. Weir, \pl{408}{97}{213};\\
D. Bellisai, F. Fucito, M. Matone and G. Travaglini, \pr{56}{97}{5218}.
\bibitem{adu} P. Argyres and M. Douglas, \np{448}{95}{93}.
\bibitem{aw} P. Argyres, M. Plesser, N. Seiberg and E. Witten, \np{461}{96}{71}.



\end{thebibliography}
\end{document}